\documentclass[3p, 11pt, sort&compress,fleqn]{elsarticle}

\usepackage{amsmath,amssymb,graphicx,siunitx,float,mathtools}
\usepackage[table]{xcolor}
\usepackage[colorlinks=true,breaklinks=true]{hyperref}
\newcommand\revision[1]{{#1}}

\makeatletter
\def\ttabular{%
\hbox\bgroup
\let\\\cr
\def\rulea{\ifnum\rowc=\@ne \hrule height 1.3pt \fi}
\def\ruleb{
\ifnum\rowc=1\hrule height 1.3pt \else
\ifnum\rowc=6\hrule height \heavyrulewidth 
   \else \hrule height \lightrulewidth\fi\fi}
\valign\bgroup
\global\rowc\@ne
\rulea
\hbox to 10em{\strut \hfill##\hfill}%
\ruleb
&&%
\global\advance\rowc\@ne
\hbox to 10em{\strut\hfill##\hfill}%
\ruleb
\cr}
\def\endttabular{%
\crcr\egroup\egroup}

\def\figw{0.47\textwidth}
\allowdisplaybreaks[4]

\makeatletter
\def\ps@pprintTitle{%
 \let\@oddhead\@empty
 \let\@evenhead\@empty
 \def\@oddfoot{}%
 \let\@evenfoot\@oddfoot}
\makeatother

\begin{document}\sloppy%\small
\begin{frontmatter}

\title{Generalized semi-analytical solution for coupled multispecies advection-dispersion equations in multilayer porous media}

\author[qut]{Elliot J. Carr}
\ead{elliot.carr@qut.edu.au}
\address[qut]{School of Mathematical Sciences, Queensland University of Technology (QUT), Brisbane, Australia.}

\journal{Applied Mathematical Modelling}

\begin{abstract}
Multispecies contaminant transport in the Earth's subsurface is commonly modelled using advection-dispersion equations coupled via first-order reactions. Analytical and semi-analytical solutions for such problems are highly sought after but currently limited to either one species, homogeneous media, certain reaction networks, specific boundary conditions or a combination thereof. In this paper, we develop a semi-analytical solution for the case of a heterogeneous layered medium and a general first-order reaction network. Our approach combines a transformation method to decouple the multispecies equations with a recently developed semi-analytical solution for the single-species advection-dispersion-reaction equation in layered media. The generalized solution is valid for arbitrary numbers of species and layers, general Robin-type conditions at the inlet and outlet and accommodates both distinct retardation factors across layers or distinct retardation factors across species. Four test cases are presented to demonstrate the solution approach with the reported results in agreement with previously published results and numerical results obtained via finite volume discretisation. MATLAB code implementing the generalized semi-analytical solution is made available.
\end{abstract}


\end{frontmatter}

\section{Introduction} 
Advection-dispersion equations are commonly used to predict the fate and transport of contaminants in the Earth's subsurface \cite{clement_2001}. Challenges of applying such equations in practical situations include dealing with subsurface heterogeneity and multiple reactive contaminants, which together yield coupled multispecies advection-dispersion-reaction equations with spatially-dependent coefficients \cite{suk_2016}. While such problems can always be solved numerically, analytical solutions are highly sought after as they provide greater insight into the governing transport processes and are useful for assessing the accuracy of numerical methods \cite{liu_1998}. Moreover, due to being continuous in space and time, analytical solutions are typically computationally efficient because computing a high accuracy solution doesn't require the use of small temporal and spatial discretisation step sizes. For such reasons, analytical solutions for reactive contaminant transport problems have attracted great interest since the mid 20th century and continue to engage researchers \cite{suk_2016,chen_2019a,chen_2019b}. 

\begin{figure*}[t]
\centering
\includegraphics[width=1.0\textwidth]{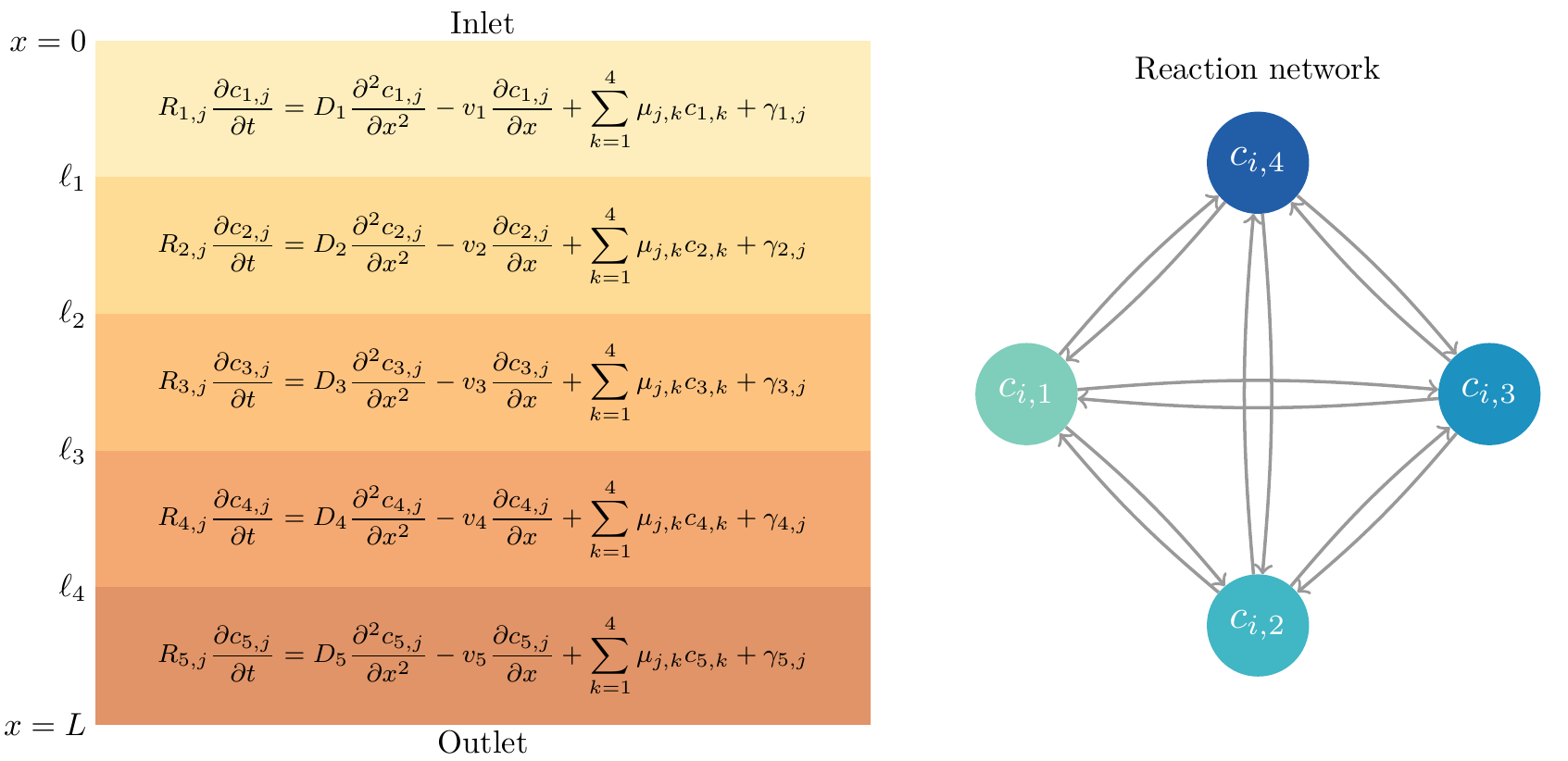}
\caption{Schematic representation of the multispecies multilayer advection-dispersion-reaction model for the case of $m = 5$ layers and $n = 4$ species, where $c_{i,j}(x,t)$ is defined as the concentration [$\text{M}\text{L}^{-3}$] of species $j$ in layer $i$. Arrows in the reaction network indicate production, e.g., the arrow from $c_{i,2}$ to $c_{i,3}$ indicates production of species 3 from species 2 in layer $i$. Our proposed solution accommodates two special cases: equal retardation factors across species ($R_{i,1} = R_{i,2} = \cdots = R_{i,n}$) with the option for distinct retardation factors across layers (Case I, Section \ref{sec:distinct_layers}) or equal retardation factors across layers ($R_{1,j} = R_{2,j} = \cdots = R_{m,j}$) with the option for distinct retardation factors across species (Case II, Section \ref{sec:distinct_species}).}
\label{fig:Figure1}
\end{figure*}

In this paper, we focus on analytical and semi-analytical solutions for reactive contaminant transport governed by coupled multispecies advection-dispersion-reaction equations. In particular, our interest is in layered media \revision{(to accommodate subsurface heterogeneity often observed in natural environments such as stratified soils or manufactured environments such as composite landfill liners \cite{liu_1998,chen_2009})} and general first order reaction networks (to accommodate each species potentially producing every other species) (Figure \ref{fig:Figure1}). In our review of the literature below, we highlight some pertinent solutions for multispecies problems in homogeneous media, single-species problems in layered media and multispecies problems in heterogeneous/layered media. 

\citet{sun_1999a} presented an analytical solution for coupled multispecies advection-dispersion-reaction equations for the case of a semi-infinite homogeneous medium with sequential reactions. Their approach involves decoupling the multispecies equations using a linear transformation and then applying a well-known analytical solution for single-species problems. The derived solution is valid for zero initial concentration, constant concentration at the inlet and a zero concentration at an infinite distance from the inlet (Figure \ref{fig:Figure1}). In subsequent work, the authors formulated other linear transformations for decoupling more complex first order reaction networks involving parallel, sequential-parallel or ``family tree'' reactions \cite{sun_1999b,sun_1999c}. Later, \citet{clement_2001} extended the analytical solutions of \citet{sun_1999a} by demonstrating how the eigenvalue decomposition of the reaction matrix (matrix of reaction rates) can be used to construct the decoupling linear transformation for a general reaction network.  

\citet{carr_2016} presented a generalized semi-analytical solution for the case of single-species diffusion in a layered medium. Similar approaches were presented by \citet{rodrigo_2016} and \citet{zimmerman_2016}; the latter for a multilayer reaction-diffusion problem with first-order decay. The approach employed by \citet{carr_2016} is to isolate the multilayer problem on each layer by introducing unknown functions equal to the diffusive flux at each interface. Each single-layer problem is then solved separately using the Laplace transform. The solution in the Laplace domain is expressed in terms of the Laplace transformations of the unknown interface functions, which are ultimately identified by ensuring the interface condition imposed on the solution is satisfied. Recently, these ideas have been extended to single-species advection-dispersion-reaction equations in layered media \cite{carr_2020}. The developed generalized solution is  valid for an arbitrary number of layers and general Robin-type boundary conditions at the inlet and outlet. 

\citet{gureghian_1985} developed a semi-analytical solution for coupled multispecies advection-dispersion-reaction equations for the case of three-species undergoing sequential reactions in a semi-infinite layered medium. The solution approach uses the Laplace transform, solving for the Laplace-domain concentration of each species in each layer in a nested fashion, which is made possible by the sequential reactions. Apart from the first layer, solutions in the time-domain are expressed in terms of convolution integrals that are evaluated numerically. \citet{suk_2013} presented an analytical solution for the case of spatially-varying retardation factor, velocity and dispersivity with sequential first-order reactions. Their approach uses the linear transformation of \citet{sun_1999a} to decouple the multispecies system with the generalized integral transform technique \cite{liu_2000} used to solve the decoupled single-species equations. The solution is valid for a time-varying flux-type inlet condition and zero concentration outlet condition. Semi-analytical solutions for multispecies multilayer problems have also been derived by \citet{mieles_2012} and \citet{chen_2018}. Both papers are limited to sequential reactions and the two-layer permeable reaction barrier-aquifer problem. 

As demonstrated by \citet{clement_2001}, the linear transformation method for decoupling multispecies problems fails for non-equal retardation factors across species. This issue was addressed by \citet{quezada_2004}, who showed how semi-analytical solutions can be obtained by applying the decoupling linear transformation in the Laplace domain. A somewhat similar approach was presented by \citet{bauer_2001} for the case of a semi-infinite domain with sequential reactions using the Laplace transform. In the Laplace domain, the solution for each species is expressed as a linear combination of decoupled single-species equations with the coefficients in the linear combination identified from a recursion formula. To transform the solution from the Laplace domain to the time domain, numerical inversion of the Laplace transform was carried out when analytical inversion was impossible. 

Our review of the literature found that all existing analytical and semi-analytical solutions for reactive contaminant transport are limited to the special cases of either a single-species, a homogeneous medium, certain reaction networks, specific boundary conditions, equal retardation factors across species or a combination thereof. In this paper, we address these limitations by developing a new generalized semi-analytical solution for multispecies contaminant transport in layered media. Our solution technique combines the decoupling strategies for multispecies \cite{clement_2001,quezada_2004} and multilayer \cite{carr_2020} advection-dispersion-reaction equations discussed previously and accommodates arbitrary numbers of species and layers, general boundary conditions of Robin-type at the inlet and outlet and a general reaction network. Seperate solutions are derived for accommodating (i) non-equal retardation factors across layers and (ii) non-equal retardation factors across species. Both solutions are implemented in MATLAB and made available in an online repository: \href{https://github.com/elliotcarr/Carr2021a}{https://github.com/elliotcarr/Carr2021a}. To the best of the author's knowledge, our generalized semi-analytical solutions are the first valid for both multilayer media and general reaction networks.

The rest of the paper is organised in the following way. In section \ref{sec:transport_model}, we outline the multispecies multilayer advection-dispersion-reaction model solved in this paper. Section \ref{sec:solution_method} describes our solution method using the decoupling transformation and Laplace transform. The derived solutions are verified in Section \ref{sec:results} using a selection of test problems. Lastly, in Section \ref{sec:conclusion}, we summarise the paper, discuss limitations of the work and provide some interesting directions for further research.

\section{Transport model}
\label{sec:transport_model}
Consider transport of a multispecies reacting contaminant within a layered medium. Let $n$ be the number of species and $m$ be the number of layers and assume the medium is layered as follows: $0 = \ell_{0} < \ell_{1} < \cdots < \ell_{m-1} < \ell_{m} = L$ (Figure \ref{fig:Figure1}). Define the concentration [$\text{M}\text{L}^{-3}$] of species $j$ in the $i$th layer by $c_{i,j}(x,t)$, where $x\in (\ell_{i-1},\ell_{i})$ and $t\geq0$ are the spatial and temporal variables. We assume transport within each layer is governed by an advection-dispersion-reaction equation, yielding the coupled system of $mn$ equations:
\begin{align}
\label{eq:c_pde1}
R_{i,j}\frac{\partial c_{i,j}}{\partial t} &= D_{i}\frac{\partial^{2}c_{i,j}}{\partial x^{2}} - v_{i}\frac{\partial c_{i,j}}{\partial x} + \sum_{k=1}^{n}\mu_{j,k}c_{i,k} + \gamma_{i,j},
\end{align}
where $x\in(\ell_{i-1},\ell_{i})$, $i = 1,\hdots,m$ (layer index) and $j = 1,\hdots,n$ (species index). Here, $R_{i,j}>0$ is the retardation factor $[-]$ in layer $i$ for species $j$, $D_{i}>0$ is the dispersion coefficient [$\text{L}^{2}\text{T}^{-1}$] in layer $i$, $v_{i}$ is the pore-water velocity [$\text{L}\text{T}^{-1}$] in layer $i$ and $\gamma_{i,j}$ is the rate constant for zero-order production [$\text{M}\text{L}^{-3}\text{T}^{-1}$] of species $j$ in layer $i$ \cite{van_genuchten_1982}. Reaction between species is governed by the values of $\mu_{j,k}$ [$\text{T}^{-1}$], which specify the rate constant for first-order decay of species $j$ ($k = j$) or the rate constant for production of species $j$ from species $k$ ($k \neq j$). 
\revision{The advection-dispersion-reaction equation (\ref{eq:c_pde1}) is valid under steady-state groundwater flow conditions, where the moisture content of water is constant in space and time \cite{van_genuchten_1982,liu_1998}.}

Initially, the concentration of each species is assumed constant within each layer:
\begin{gather}
\label{eq:c_int}
c_{i,j}(x,0) = f_{i,j},
\end{gather}
where $i = 1,\hdots,m$ and $j = 1,\hdots,n$. For all species, concentration and flux are assumed continuous at the interfaces between adjacent layers \cite{leij_1991,liu_1998,guerrero_2013}:
\begin{gather}
\label{eq:c_ic1}
c_{i,j}(\ell_{i},t) = c_{i+1,j}(\ell_{i},t),\\
\label{eq:c_ic2}
\theta_{i}D_{i}\frac{\partial c_{i,j}}{\partial x}(\ell_{i},t) = \theta_{i+1}D_{i+1}\frac{\partial c_{i+1,j}}{\partial x}(\ell_{i},t),
\end{gather}
where $i = 1,\hdots,m-1$ and $j = 1,\hdots,n$ and $\theta_{i}$ is the volumetric water content $[-]$ in layer $i$. Inherent in Eq (\ref{eq:c_ic2}) is the assumption of flow continuity $\theta_{i}v_{i} = \theta_{i+1}v_{i+1}$ for all $i = 1,\hdots,m-1$ \cite{leij_1991,liu_1998}. General Robin boundary conditions are applied at the inlet and outlet:
\begin{gather}
\label{eq:c_bc1}
a_{0}c_{1,j}(0,t) - b_{0}\frac{\partial c_{1,j}}{\partial x}(0,t) = g_{0,j}(t),\\
\label{eq:c_bc2}
a_{L}c_{m,j}(L,t) + b_{L}\frac{\partial c_{m,j}}{\partial x}(L,t) = g_{m,j}(t),
\end{gather}
where $j = 1,\hdots,n$, $a_{0}, b_{0}, a_{L}, b_{L} > 0$ are specified constants and $g_{0,j}(t)$ and $g_{m,j}(t)$ are specified functions. The Robin boundary conditions  (\ref{eq:c_bc1})--(\ref{eq:c_bc2}) permit either a concentration-type boundary condition, for example by setting $a_{0} = 1$, $b_{0} = 0$ and $g_{0,j}(t) = c_{b,j}(t)$, or a flux-type boundary condition, for example by setting $a_{0} = v_{1}$, $b_{0} = D_{1}$ and $g_{0,j}(t) = v_{1}c_{b,j}(t)$, where $c_{b,j}(t)$ is the specified inlet concentration of species $j$.

\section{Solution method}
\label{sec:solution_method}
Our generalized semi-analytical method for solving the coupled multispecies multilayer transport model (\ref{eq:c_pde1})--(\ref{eq:c_bc2}) is now described. We start by considering the following matrix representation of the governing equations (\ref{eq:c_pde1})--(\ref{eq:c_bc2}):
\begin{gather}
\label{eq:c_pde_mf}
\mathbf{R}_{i}\frac{\partial\mathbf{c}_{i}}{\partial t} = D_{i}\frac{\partial^{2}\mathbf{c}_{i}}{\partial x^{2}} - v_{i}\frac{\partial\mathbf{c}_{i}}{\partial x} + \mathbf{M}\mathbf{c}_{i} + \boldsymbol{\gamma}_{i},\quad\text{(for $x\in(\ell_{i-1},\ell_{i})$ and $i = 1,\hdots,m$)},\\
\label{eq:c_ic_mf}
\mathbf{c}_{i}(x,0) = \mathbf{f}_{i},\quad\text{(for $x\in(\ell_{i-1},\ell_{i})$ and $i = 1,\hdots,m$)},\\
%\nonumber \text{(for $x\in(\ell_{i-1},\ell_{i})$ and $i = 1,\hdots,m$)},\\
\mathbf{c}_{i}(\ell_{i},t) = \mathbf{c}_{i+1}(\ell_{i},t),\quad\text{(for $i = 1,\hdots,m-1$)},\\
\label{eq:c_ic2_mf}
\theta_{i}D_{i}\frac{\partial\mathbf{c}_{i}}{\partial x}(\ell_{i},t) = \theta_{i+1}D_{i+1}\frac{\partial\mathbf{c}_{i+1}}{\partial x}(\ell_{i},t),\quad\text{(for $i = 1,\hdots,m-1$)},\\
%\nonumber \text{(for $i = 1,\hdots,m-1$)},\\
\label{eq:c_bc1_mf}
a_{0}\mathbf{c}_{1}(0,t) - b_{0}\frac{\partial\mathbf{c}_{1}}{\partial x}(0,t) = \mathbf{g}_{0}(t),\\
\label{eq:c_bc2_mf}
a_{L}\mathbf{c}_{m}(L,t) + b_{L}\frac{\partial\mathbf{c}_{m}}{\partial x}(L,t) = \mathbf{g}_{m}(t),
\end{gather}
where the vectors and matrices are defined as follows:
\begin{gather*}
\mathbf{c}_{i} = \left[\begin{matrix} c_{i,1}\\ c_{i,2}\\ \vdots\\ c_{i,n}\end{matrix}\right],\quad \mathbf{R}_{i} = \left[\begin{matrix} R_{i,1} &  &  & \\ & R_{i,2} & & \\ & & \ddots & \\ & & & R_{i,n}\end{matrix}\right],\quad 
\mathbf{M}_{i} = \left[\begin{matrix} \mu_{1,1} & \mu_{1,2} & \cdots & \mu_{1,n}\\ \mu_{2,1} & \mu_{2,2} & \cdots & \mu_{2,n}\\ \vdots & \vdots & \ddots & \vdots\\ \mu_{n,1} & \mu_{n,2} & \cdots & \mu_{n,n}\end{matrix}\right],\\ \boldsymbol{\gamma}_{i} = \left[\begin{matrix} \gamma_{i,1}\\ \gamma_{i,2}\\ \vdots\\ \gamma_{i,n}\end{matrix}\right],\quad \mathbf{f}_{i} = \left[\begin{matrix} f_{i,1}\\ f_{i,2}\\ \vdots\\ f_{i,n}\end{matrix}\right],\quad \mathbf{g}_{0}(t) = \left[\begin{matrix} g_{0,1}(t)\\ g_{0,2}(t)\\ \vdots\\ g_{0,n}(t)\end{matrix}\right],\quad \mathbf{g}_{m}(t) = \left[\begin{matrix} g_{m,1}(t)\\ g_{m,2}(t)\\ \vdots\\ g_{m,n}(t)\end{matrix}\right].
\end{gather*}
Our proposed solution accommodates two special cases: equal retardation factors across species with the option for non-equal retardation factors across layers (Case I) or equal retardation factors across layers with the option for non-equal retardation factors across species (Case II). Below, we consider these two cases separately.

\subsection{Case I: Non-equal retardation factors across layers}
\label{sec:distinct_layers}
Suppose within each layer the retardation factors are equal for each species, that is, $R_{i,1} = R_{i,2} = \cdots = R_{i,n} =: \widehat{R}_{i}$, where $\widehat{R}_{i}$ is the retardation factor for all species in layer $i$. In this case, $\mathbf{R}_{i} = \widehat{R}_{i}\mathbf{I}$ and the coupled system of transport equations in layer $i$ (\ref{eq:c_pde_mf}) becomes:
\begin{gather}
\label{eq:c_pde_mf_CaseI}
\widehat{R}_{i}\frac{\partial\mathbf{c}_{i}}{\partial t} = D_{i}\frac{\partial^{2}\mathbf{c}_{i}}{\partial x^{2}} - v_{i}\frac{\partial\mathbf{c}_{i}}{\partial x} + \mathbf{M}\mathbf{c}_{i} + \boldsymbol{\gamma}_{i}.
\end{gather}
To decouple this system of $n$ equations, we employ the linear transformation approach of \citet{clement_2001}. Let $\widetilde{\mathbf{c}}_{i} = \mathbf{Q}^{-1}\mathbf{c}_{i}$ where $\mathbf{Q}$ is an $n\times n$ invertible matrix. Multiplying (\ref{eq:c_pde_mf_CaseI}) by $\mathbf{Q}^{-1}$ yields:
\begin{align*}
\widehat{R}_{i}\frac{\partial\widetilde{\mathbf{c}}_{i}}{\partial t} &= D_{i}\frac{\partial^{2}\widetilde{\mathbf{c}}_{i}}{\partial x^{2}} - v_{i}\frac{\partial\widetilde{\mathbf{c}}_{i}}{\partial x} + \left(\mathbf{Q}^{-1}\mathbf{M}\mathbf{Q}\right)\widetilde{\mathbf{c}}_{i} + \mathbf{Q}^{-1}\boldsymbol{\gamma}_{i},
\end{align*}
where we have used the linearity of the differential operator and the relationships between $\mathbf{c}_{i}$ and $\widetilde{\mathbf{c}}_{i}$. Hence, the coupling between species can be removed by choosing $\mathbf{Q}$ such that $\mathbf{Q}^{-1}\mathbf{M}\mathbf{Q}$ is a diagonal matrix. Assuming $\mathbf{M}$ is non-defective (diagonalisable), an appropriate choice for $\mathbf{Q}$ is obtained by computing the eigenvalue decomposition: $\mathbf{M} = \mathbf{Q}\boldsymbol{\Lambda}\mathbf{Q}^{-1}$, where the diagonal entries of $\boldsymbol{\Lambda} = \mathbf{Q}^{-1}\mathbf{M}\mathbf{Q} = \text{diag}(\lambda_{1},\lambda_{2},\hdots,\lambda_{n})$ are the eigenvalues of $\mathbf{M}$ and the columns of $\mathbf{Q}$ are the corresponding eigenvectors \cite{clement_2001}. 

All there is left to do is to transform the initial, boundary and interface conditions (\ref{eq:c_ic_mf})--(\ref{eq:c_bc2_mf}), which is achieved by multiplying both sides of each of these equations by $\mathbf{Q}^{-1}$. In summary, solving (\ref{eq:c_pde_mf})--(\ref{eq:c_ic2_mf}) for $\mathbf{c}_{i}$ is equivalent to solving:
\begin{gather}
\label{eq:ct_pde_mf}
\widehat{R}_{i}\frac{\partial\widetilde{\mathbf{c}}_{i}}{\partial t} = D_{i}\frac{\partial^{2}\widetilde{\mathbf{c}}_{i}}{\partial x^{2}} - v_{i}\frac{\partial\widetilde{\mathbf{c}}_{i}}{\partial x} + \boldsymbol{\Lambda}\widetilde{\mathbf{c}}_{i} + \widetilde{\boldsymbol{\gamma}}_{i},\quad\text{(for $x\in(\ell_{i-1},\ell_{i})$ and $i = 1,\hdots,m$)},\\
\label{eq:ct_ic_mf}
\widetilde{\mathbf{c}}_{i}(x,0) = \widetilde{\mathbf{f}}_{i},\quad\text{(for $x\in(\ell_{i-1},\ell_{i})$ and $i = 1,\hdots,m$)},\\
%\nonumber \text{(for $x\in(\ell_{i-1},\ell_{i})$ and $i = 1,\hdots,m$)},\\
\label{eq:ct_ic1_mf}
\widetilde{\mathbf{c}}_{i}(\ell_{i},t) = \widetilde{\mathbf{c}}_{i+1}(\ell_{i},t),\quad\text{(for $i = 1,\hdots,m-1$)},\\
\label{eq:ct_ic2_mf}
\theta_{i}D_{i}\frac{\partial\widetilde{\mathbf{c}}_{i}}{\partial x}(\ell_{i},t) = \theta_{i+1}D_{i+1}\frac{\partial\widetilde{\mathbf{c}}_{i+1}}{\partial x}(\ell_{i},t),\quad\text{(for $i = 1,\hdots,m-1$)},\\
%\nonumber \text{(for $i = 1,\hdots,m-1$)},\\
\label{eq:ct_bc1_mf}
a_{0}\widetilde{\mathbf{c}}_{1}(0,t) - b_{0}\frac{\partial\widetilde{\mathbf{c}}_{1}}{\partial x}(0,t) = \widetilde{\mathbf{g}}_{0}(t),\\
\label{eq:ct_bc2_mf}
a_{L}\widetilde{\mathbf{c}}_{m}(L,t) + b_{L}\frac{\partial\widetilde{\mathbf{c}}_{m}}{\partial x}(L,t) = \widetilde{\mathbf{g}}_{m}(t),
\end{gather}
for $\widetilde{\mathbf{c}}_{i}$ and then computing $\mathbf{c}_{i} = \mathbf{Q}\widetilde{\mathbf{c}}_{i}$. In the above equations, $\widetilde{\boldsymbol{\gamma}}_{i}=\mathbf{Q}^{-1}\boldsymbol{\gamma}_{i}$, $\widetilde{\mathbf{f}}_{i}=\mathbf{Q}^{-1}\mathbf{f}_{i}$, $\widetilde{\mathbf{g}}_{0}(t) = \mathbf{Q}^{-1}\mathbf{g}_{0}(t)$ and $\widetilde{\mathbf{g}}_{m}(t) = \mathbf{Q}^{-1}\mathbf{g}_{m}(t)$.

Collecting the $j$th component of the system of equations (\ref{eq:ct_pde_mf})--(\ref{eq:ct_bc2_mf}) for each $i = 1,\hdots,m$, provides the decoupled problem for species $j$: 
\begin{gather}
\label{eq:ct_pde_j}
\widehat{R}_{i}\frac{\partial \widetilde{c}_{i,j}}{\partial t} = D_{i}\frac{\partial^{2}\widetilde{c}_{i,j}}{\partial x^{2}} - v_{i}\frac{\partial \widetilde{c}_{i,j}}{\partial x} + \lambda_{j}\widetilde{c}_{i,j} + \widetilde{\gamma}_{i,j},\quad\text{(for $x\in(\ell_{i-1},\ell_{i})$ and $i = 1,\hdots,m$)},\\
\label{eq:ct_ic_j}
\widetilde{c}_{i,j}(x,0) = \widetilde{f}_{i,j},\quad\text{(for $x\in(\ell_{i-1},\ell_{i})$ and $i = 1,\hdots,m$)},\\
%\nonumber \text{(for $x\in(\ell_{i-1},\ell_{i})$ and $i = 1,\hdots,m$)},\\
\label{eq:ct_ic1_j}
\widetilde{c}_{i,j}(\ell_{i},t) = \widetilde{c}_{i+1,j}(\ell_{i},t),\quad\text{(for $i = 1,\hdots,m-1$)},\\
\label{eq:ct_ic2_j}
\theta_{i}D_{i}\frac{\partial \widetilde{c}_{i,j}}{\partial x}(\ell_{i},t) = \theta_{i+1}D_{i+1}\frac{\partial \widetilde{c}_{i+1,j}}{\partial x}(\ell_{i},t),\quad\text{(for $i = 1,\hdots,m-1$)},\\
%\nonumber \text{(for $i = 1,\hdots,m-1$)},\\
\label{eq:ct_bc1_j}
a_{0}\widetilde{c}_{1,j}(0,t) - b_{0}\frac{\partial \widetilde{c}_{1,j}}{\partial x}(0,t) = \widetilde{g}_{0,j}(t),\\
\label{eq:ct_bc2_j}
a_{L}\widetilde{c}_{m,j}(L,t) + b_{L}\frac{\partial \widetilde{c}_{m,j}}{\partial x}(L,t) = \widetilde{g}_{m,j}(t),
\end{gather}
where $\widetilde{f}_{i,j}$, $\widetilde{\gamma}_{i,j}$, $\widetilde{g}_{0,j}(t)$ and $\widetilde{g}_{m,j}(t)$ are the $j$th entries of $\widetilde{\mathbf{f}}_{i}$, $\widetilde{\boldsymbol{\gamma}}_{i}$, $\widetilde{\mathbf{g}}_{0}(t)$ and $\widetilde{\mathbf{g}}_{m}(t)$, respectively. For each value of $j$, the governing equations (\ref{eq:ct_pde_j})--(\ref{eq:ct_bc2_j}) describe a single-species multilayer advection-dispersion-reaction model. We solve this model using the semi-analytical method presented in our previous work \cite{carr_2020}, which can be applied directly to (\ref{eq:ct_pde_j})--(\ref{eq:ct_bc2_j}) for each species $j = 1,\hdots,n$. Using the computed solution $\widetilde{c}_{i,j}(x,t)$ and reversing the decoupling transformation yields the solution of the original coupled multispecies multilayer transport model: $c_{i,j}(x,t) = \sum_{k=1}^{n}q_{j,k}\widetilde{c}_{i,k}(x,t)$, where $q_{j,k}$ is the entry in row $j$ and column $k$ of $\mathbf{Q}$.

\subsection{Case II: Non-equal retardation factors across species}
\label{sec:distinct_species}
Suppose for each species the retardation factors are equal within each layer, that is, $R_{1,j} = R_{2,j} = \cdots = R_{m,j} =: \overline{R}_{j}$, where $\overline{R}_{j}$ is the retardation factor for species $j$ across all layers. In this case, $\mathbf{R}_{i} = \text{diag}(\overline{R}_{1},\overline{R}_{2},\hdots,\overline{R}_{n}) =: \overline{\mathbf{R}}$ and the coupled system of transport equations in layer $i$ (\ref{eq:c_pde_mf}) becomes:
\begin{gather}
\label{eq:c_pde_mf_CaseII}
\overline{\mathbf{R}}\frac{\partial\mathbf{c}_{i}}{\partial t} = D_{i}\frac{\partial^{2}\mathbf{c}_{i}}{\partial x^{2}} - v_{i}\frac{\partial\mathbf{c}_{i}}{\partial x} + \mathbf{M}\mathbf{c}_{i} + \boldsymbol{\gamma}_{i}.
\end{gather}
Attempting to decouple this system using the strategy outlined for Case I fails since $\mathbf{Q}^{-1}\overline{\mathbf{R}}[\partial\mathbf{c}_{i}/\partial t] = \overline{\mathbf{R}}[\partial\widetilde{\mathbf{c}}_{i}/\partial t]$ only if $\mathbf{Q}^{-1}\overline{\mathbf{R}} = \overline{\mathbf{R}}\mathbf{Q}^{-1}$, i.e., $\overline{\mathbf{R}}$ is diagonal with equal diagonal entries (as in Case I) \cite{clement_2001}. However, non-equal retardation factors across species can be accommodated by applying the linear transformation in the Laplace domain \cite{quezada_2004}. Consider the Laplace domain representation of Eqs (\ref{eq:c_pde_mf_CaseII}) and (\ref{eq:c_ic_mf})--(\ref{eq:c_bc2_mf}):
\begin{gather}
\label{eq:C_de_mf}
\overline{\mathbf{R}}\left[s\mathbf{C}_{i} - \mathbf{f}_{i}\right] = D_{i}\frac{\text{d}^{2}\mathbf{C}_{i}}{\text{d}x^{2}} - v_{i}\frac{\text{d}\mathbf{C}_{i}}{\text{d}x} + \mathbf{M}\mathbf{C}_{i} + \frac{\boldsymbol{\gamma}_{i}}{s},\quad\text{(for $x\in(\ell_{i-1},\ell_{i})$ and $i = 1,\hdots,m$)},\\
%\nonumber \text{(for $x\in(\ell_{i-1},\ell_{i})$ and $i = 1,\hdots,m$)},\\
\label{eq:C_ic1_mf}
\mathbf{C}_{i}(\ell_{i},s) = \mathbf{C}_{i+1}(\ell_{i},s),\quad\text{(for $i = 1,\hdots,m-1$)},\\
\label{eq:C_ic2_mf}
\theta_{i}D_{i}\frac{\partial\mathbf{C}_{i}}{\partial x}(\ell_{i},s) = \theta_{i+1}D_{i+1}\frac{\partial\mathbf{C}_{i+1}}{\partial x}(\ell_{i},s),\quad\text{(for $i = 1,\hdots,m-1$)},\\
%\nonumber \text{(for $i = 1,\hdots,m-1$)},\\
\label{eq:C_bc1_mf}
a_{0}\mathbf{C}_{1}(0,s) - b_{0}\frac{\partial\mathbf{C}_{1}}{\partial x}(0,s) = \mathbf{G}_{0}(s),\\
\label{eq:C_bc2_mf}
a_{L}\mathbf{C}_{m}(L,s) + b_{L}\frac{\partial\mathbf{C}_{m}}{\partial x}(L,s) = \mathbf{G}_{m}(s),
\end{gather}
where $\mathbf{C}_{i} = \left[C_{i,1},C_{i,2},\hdots,C_{i,n}\right]^{T}$ with $C_{i,j} := C_{i,j}(x,s) = \mathcal{L}\{c_{i,j}(x,t)\}$. Let $\widetilde{\mathbf{C}}_{i} = \mathbf{Q}^{-1}\mathbf{C}_{i}$, where $\mathbf{Q}$ is an $n\times n$ invertible  matrix. Rearranging (\ref{eq:C_de_mf}) and multiplying both sides of the resulting rearrangement by $\mathbf{Q}^{-1}$ yields:
\begin{gather*}
\mathbf{Q}^{-1}\left[-\overline{\mathbf{R}}\mathbf{f}_{i}-\frac{\boldsymbol{\gamma}_{i}}{s}\right] = D_{i}\frac{\text{d}^{2}\widetilde{\mathbf{C}}_{i}}{\text{d}x^{2}} - v_{i}\frac{\text{d}\widetilde{\mathbf{C}}_{i}}{\text{d}x} + \left(\mathbf{Q}^{-1}\left[\mathbf{M}-s\overline{\mathbf{R}}\right]\mathbf{Q}\right)\widetilde{\mathbf{C}}_{i},
\end{gather*}
where, similarly to Case I, we have used the linearity of the differential operator and the relationships between $\mathbf{C}_{i}$ and $\widetilde{\mathbf{C}}_{i}$. This time assuming $\mathbf{M}-s\overline{\mathbf{R}}$ is non-defective, the coupling between species in the Laplace domain can be removed by computing the eigenvalue decomposition: $\mathbf{M}-s\overline{\mathbf{R}} = \mathbf{Q}\boldsymbol{\Lambda}\mathbf{Q}^{-1}$  where the diagonal entries of $\boldsymbol{\Lambda} = \text{diag}(\lambda_{1},\lambda_{2},\hdots,\lambda_{n})$ are the eigenvalues $\mathbf{M}-s\overline{\mathbf{R}}$ and the columns of $\mathbf{Q}$ are the corresponding eigenvectors \cite{quezada_2004}. Clearly, the entries of $\mathbf{Q}$ and $\boldsymbol{\Lambda}$ depend on the Laplace variable, $s$, so henceforth we write $\mathbf{Q}(s)$ and $\boldsymbol{\Lambda}(s) = \text{diag}(\lambda_{1}(s),\lambda_{2}(s),\hdots,\lambda_{n}(s))$ to explicitly highlight this dependence. In a similar manner to Case I, transforming the boundary and interface conditions (\ref{eq:C_ic1_mf})--(\ref{eq:C_bc2_mf}) is achieved by multiplying by $[\mathbf{Q}(s)]^{-1}$. In summary, $\mathbf{\widetilde{C}}_{i}$ satisfies:
\begin{gather}
\label{eq:Ct_de_mf}
\boldsymbol{\alpha}_{i}(s) + \frac{\boldsymbol{\omega}_{i}(s)}{s} = D_{i}\frac{\text{d}^{2}\widetilde{\mathbf{C}}_{i}}{\text{d}x^{2}} - v_{i}\frac{\text{d}\widetilde{\mathbf{C}}_{i}}{\text{d}x} + \boldsymbol{\Lambda}(s)\widetilde{\mathbf{C}}_{i},\quad\text{(for $x\in(\ell_{i-1},\ell_{i})$ and $i = 1,\hdots,m$)},\\
%\nonumber \text{(for $x\in(\ell_{i-1},\ell_{i})$ and $i = 1,\hdots,m$)},\\
\label{eq:Ct_ic1_mf}
\widetilde{\mathbf{C}}_{i}(\ell_{i},s) = \widetilde{\mathbf{C}}_{i+1}(\ell_{i},s),\quad\text{(for $i = 1,\hdots,m-1$),}\\
\label{eq:Ct_ic2_mf}
\theta_{i}D_{i}\frac{\partial\widetilde{\mathbf{C}}_{i}}{\partial x}(\ell_{i},s) = \theta_{i+1}D_{i+1}\frac{\partial\widetilde{\mathbf{C}}_{i+1}}{\partial x}(\ell_{i},s),\quad\text{(for $i = 1,\hdots,m-1$),}\\
%\nonumber \text{(for $i = 1,\hdots,m-1$),}\\
\label{eq:Ct_bc1_mf}
a_{0}\widetilde{\mathbf{C}}_{1}(0,s) - b_{0}\frac{\partial\widetilde{\mathbf{C}}_{1}}{\partial x}(0,s) = \widetilde{\mathbf{G}}_{0}(s),\\
\label{eq:Ct_bc2_mf}
a_{L}\widetilde{\mathbf{C}}_{m}(L,s) + b_{L}\frac{\partial\widetilde{\mathbf{C}}_{m}}{\partial x}(L,s) = \widetilde{\mathbf{G}}_{m}(s),
\end{gather}
%$\mathbf{C}_{i} = \mathbf{Q}(s)\widetilde{\mathbf{C}}_{i}$. 
where $\boldsymbol{\alpha}_{i}(s) = -[\mathbf{Q}(s)]^{-1}\overline{\mathbf{R}}\mathbf{f}_{i}$, $\boldsymbol{\omega}_{i}(s) = -[\mathbf{Q}(s)]^{-1}\boldsymbol{\gamma}_{i}$, $\widetilde{\mathbf{G}}_{0}(s) = [\mathbf{Q}(s)]^{-1}\mathbf{G}_{0}(s)$ and $\widetilde{\mathbf{G}}_{m}(s) = [\mathbf{Q}(s)]^{-1}\mathbf{G}_{m}(s)$.

Collecting the $j$th component of the system of equations (\ref{eq:Ct_de_mf})--(\ref{eq:Ct_bc2_mf}) for each $i = 1,\hdots,m$, provides the decoupled problem for species $j$: 
\begin{gather}
\label{eq:Ctj_pde}
\alpha_{i,j}(s) + \frac{\omega_{i,j}(s)}{s} = D_{i}\frac{\text{d}^{2}\widetilde{C}_{i,j}}{\text{d}x^{2}} - v_{i}\frac{\text{d}\widetilde{C}_{i,j}}{\text{d}x} + \lambda_{j}(s)\widetilde{C}_{i,j},\quad\text{(for $x\in(\ell_{i-1},\ell_{i})$ and $i = 1,\hdots,m$),}\\
%\nonumber ,\\
\label{eq:Ctj_ic1}
\widetilde{C}_{i,j}(\ell_{i},s) = \widetilde{C}_{i+1,j}(\ell_{i},s),\quad\text{(for $i = 1,\hdots,m-1$),}\\
\label{eq:Ctj_ic2}
\theta_{i}D_{i}\frac{\partial\widetilde{C}_{i,j}}{\partial x}(\ell_{i},s) = \theta_{i+1}D_{i+1}\frac{\partial\widetilde{C}_{i+1,j}}{\partial x}(\ell_{i},s),\quad\text{(for $i = 1,\hdots,m-1$),}\\
%\nonumber \\
\label{eq:Ctj_bc1}
a_{0}\widetilde{C}_{1,j}(0,s) - b_{0}\frac{\partial\widetilde{C}_{1,j}}{\partial x}(0,s) = \widetilde{G}_{0,j}(s),\\
\label{eq:Ctj_bc2}
a_{L}\widetilde{C}_{m,j}(L,s) + b_{L}\frac{\partial\widetilde{C}_{m,j}}{\partial x}(L,s) = \widetilde{G}_{m,j}(s),
\end{gather}
where $\alpha_{i,j}(s)$, $\omega_{i,j}(s)$, $\widetilde{G}_{0,j}(s)$, $\widetilde{G}_{m,j}(s)$ denote the $j$th component of the vectors $\boldsymbol{\alpha}_{i}(s)$, $\boldsymbol{\omega}_{i}(s)$, $\widetilde{\mathbf{G}}_{0}(s)$ and $\widetilde{\mathbf{G}}_{m}(s)$.

Following similar working to that presented previously by \citet{carr_2020}, the solution of Eqs (\ref{eq:Ctj_pde}) and (\ref{eq:Ctj_ic2})--(\ref{eq:Ctj_bc2}) can be expressed as:
\begin{align}
\label{eq:C1}
\widetilde{C}_{1,j}(x,s) &= P_{1,j}(x,s) + A_{1,j}(x,s)\widetilde{G}_{0,j}(s) + B_{1,j}(x,s)\widetilde{G}_{1,j}(s),\\
\label{eq:Ci}
\widetilde{C}_{i,j}(x,s) &= P_{i,j}(x,s) + A_{i,j}(x,s)\widetilde{G}_{i-1,j}(s) + B_{i,j}(x,s)\widetilde{G}_{i,j}(s),\quad\text{(for $i = 2,\hdots,m-1$),}\\
%\nonumber &\text{(for $i = 2,\hdots,m-1$),}\\
\label{eq:Cm}
\widetilde{C}_{m,j}(x,s) &= P_{m,j}(x,s) + A_{m,j}(x,s)\widetilde{G}_{m-1}(s) + B_{m,j}(x,s)\widetilde{G}_{m,j}(s),
\end{align}
where $P_{i,j}$, $A_{i,j}$ and $B_{i,j}$ ($i = 1,\hdots,m$) are defined in Table \ref{tab:FAB} and $\widetilde{G}_{i,j}(s) := \theta_{i}D_{i}[\partial\widetilde{C}_{i,j}/\partial x](\ell_{i},s)$ for $i = 1,\hdots,m-1$ and $j = 1,\hdots,n$. Substituting the expressions (\ref{eq:C1})--(\ref{eq:Cm}) into the interface condition (\ref{eq:Ctj_ic1}) for $i = 1,\hdots,m-1$ yields an $(m-1)$-dimensional linear system whose solution provides $\widetilde{G}_{1,j}(s),\hdots,\widetilde{G}_{m-1,j}$ for species $j$. With $\widetilde{C}_{i,j}(x,s)$ identified in all layers, reversing the decoupling transformation yields $C_{i,j}(x,s) = \sum_{k=1}^{n}q_{j,k}(s)\widetilde{C}_{i,k}(x,s)$, where $q_{j,k}(s)$ is the entry in row $j$ and column $k$ of $\mathbf{Q}(s)$. 

\revision{The final step is to apply the inverse Laplace transform to compute the concentration in the time domain. Here, we use an approach described by \citet{trefethen_2006} to numerically invert the Laplace transform:
\begin{gather}
\label{eq:num_inv_lap}
c_{i,j}(x,t) = \mathcal{L}^{-1}\left\{C_{i,j}(x,s)\right\} = -\frac{2}{t}\Re\left\{\sum_{k\in O_{N}}w_{k}C_{i,j}(x,s_{k})\right\},
\end{gather}
where $s_{k} = z_{k}/t$, $N$ is even, $O_{N}$ is the set of positive odd integers less than $N$ and $w_{k}$ and $z_{k}$ are complex constants denoting the residues and poles of the best $(N,N)$ rational approximation to $e^{z}$ on the negative real line ($\Re(z) < 0$). This quadrature formula is derived in \cite{trefethen_2006} from the contour integral representation of the inverse Laplace transform by inserting the aforementioned best rational approximant and then using residue calculus. Full implementation details are available in our code available at \href{https://github.com/elliotcarr/Carr2021a}{https://github.com/elliotcarr/Carr2021a}.

Trefethen's \cite{trefethen_2006} method is very similar in terms of form and computational efficiency to the Zakian method \cite{zakian_1969}. Similarly to that reported by \citet{wang_2015} for the Zakian method, we found that (\ref{eq:num_inv_lap}) can also lead to unreliable results for advection-dominated problems.  For such problems it may be beneficial to investigate other methods of numerical Laplace transform inversion recommended in \cite{wang_2015} for advection-dominated problems, such as the Simon, Weeks or de Hoog methods \cite{wang_2015}.} 

\begin{table*}[p]
\centering
\caption{Form of $P_{i,j}(x,s)$, $A_{i,j}(x,s)$ and $B_{i,j}(x,s)$ ($i = 1,\hdots,m$) involved in Eqs (\ref{eq:C1})--(\ref{eq:Cm}).}
\renewcommand{\arraystretch}{1.6}
\begin{tabular}{l}
\hline
\textbf{Layer adjacent to inlet} ($i = 1$)\\ 
$\displaystyle P_{1,j}(x,s) = \Psi_{1}^{(j)}(s) + \frac{a_{0}}{\beta_{1,j}(s)}\Bigl\{\xi_{1,1}^{(j)}(s)\Psi_{1,2}^{(j)}(x,s) - \xi_{1,2}^{(j)}(s)\Psi_{1,2}^{(j)}(\ell_{1},s)\Psi_{1,1}^{(j)}(x,s)\Bigr\}\Psi_{1}^{(j)}(s)$,\\
$\displaystyle A_{1,j}(x,s) = \frac{1}{\beta_{1,j}(s)}\Bigl\{\xi_{1,2}^{(j)}(s)\Psi_{1,2}^{(j)}(\ell_{1},s)\Psi_{1,1}^{(j)}(x,s) - \xi_{1,1}^{(j)}(s)\Psi_{1,2}^{(j)}(x,s)\Bigr\},$\\
$\displaystyle B_{1,j}(x,s) = \frac{1}{\theta_{1}D_{1}\beta_{1,j}(s)}\Bigl\{\bigl[a_{0}-b_{0}\xi_{1,1}^{(j)}(s)\bigr]\Psi_{1,1}^{(j)}(0,s)\Psi_{1,2}^{(j)}(x,s) - \bigl[a_{0} - b_{0}\xi_{1,2}^{(j)}(s)\bigr]\Psi_{1,1}^{(j)}(x,s)\Bigr\}$,\\
\text{where}\\
$\displaystyle \beta_{1,j}(s) = [a_{0} - b_{0}\xi_{1,1}^{(j)}(s)]\xi_{1,2}^{(j)}(s)\exp(-[\xi_{1,1}^{(j)}(s)-\xi_{1,2}^{(j)}(s)]\ell_{1}) - [a_{0} - b_{0}\xi_{1,2}^{(j)}(s)]\xi_{1,1}^{(j)}(s)$.\\
\hline
\textbf{Interior layers} ($i = 2,\hdots,m-1$)\\
$\displaystyle P_{i,j}(x,s) = \Psi_{i}^{(j)}(s)$,\\
$\displaystyle A_{i,j}(x,s) = \frac{1}{\theta_{i}D_{i}\beta_{i,j}(s)}\Bigl\{\xi_{i,2}^{(j)}(s)\Psi_{i,2}^{(j)}(\ell_{i},s)\Psi_{i,1}^{(j)}(x,s) - \xi_{i,1}^{(j)}(s)\Psi_{i,2}^{(j)}(x,s)\Bigr]$,\\
$\displaystyle B_{i,j}(x,s) = \frac{1}{\theta_{i}D_{i}\beta_{i,j}(s)}\Bigl\{\xi_{i,1}^{(j)}(s)\Psi_{i,1}^{(j)}(\ell_{i-1},s)\Psi_{i,2}^{(j)}(x,s)-\xi_{i,2}^{(j)}(s)\Psi_{i,1}^{(j)}(x,s)\Bigr\}$,\\
\text{where}\\
$\displaystyle \beta_{i,j}(s) = \xi_{i,1}^{(j)}(s)\xi_{i,2}^{(j)}(s)\left\{\exp(-[\xi_{i,1}^{(j)}(s)-\xi_{i,2}^{(j)}(s)](\ell_{i}-\ell_{i-1}))-1\right\}$.\\
\hline
\textbf{Layer adjacent to outlet} ($i = m$)\\
$\displaystyle P_{m,j}(x,s) = \Psi_{m}^{(j)}(s) + \frac{a_{L}}{\beta_{m,j}(s)}\Bigl\{\xi_{m,2}^{(j)}(s)\Psi_{m,1}^{(j)}(x,s) - \xi_{m,1}^{(j)}(s)\Psi_{m,1}^{(j)}(\ell_{m-1},s)\Psi_{m,2}^{(j)}(x,s)\Bigr\}\Psi_{m,1}^{(j)}(s)$,\\
$\displaystyle A_{m,j}(x,s) = \frac{1}{\theta_{m}D_{m}\beta_{m,j}(s)}\Bigl\{\bigl[a_{L}+b_{L}\xi_{m,2}^{(j)}(s)\bigr]\Psi_{m,2}^{(j)}(L,s)\Psi_{m,1}^{(j)}(x,s) - \left[a_{L}+b_{L}\xi_{m,1}^{(j)}(s)\right]\Psi_{m,2}^{(j)}(x,s)\Bigr\}$,\\
$\displaystyle B_{m,j}(x,s) = \frac{1}{\beta_{m,j}(s)}\Bigl\{\xi_{m,1}^{(j)}(s)\Psi_{m,1}^{(j)}(\ell_{m-1},s)\Psi_{m,2}^{(j)}(x,s) - \xi_{m,2}^{(j)}(s)\Psi_{m,1}^{(j)}(x,s)\Bigr\}$,\\
\text{where}\\
$\displaystyle \beta_{m,j}(s) = [a_{L}+b_{L}\xi_{m,2}^{(j)}(s)]\xi_{m,1}^{(j)}(s)\exp(-[\xi_{m,1}^{(j)}(s)-\xi_{m,2}^{(j)}(s)](\ell_{m}-\ell_{m-1})) - [a_{L}+b_{L}\xi_{m,1}^{(j)}(s)]\xi_{m,2}^{(j)}(s)$.\\
\hline
For every species ($j = 1,\hdots,n$) and every layer ($i = 1,\hdots,m$):\\
$\displaystyle\Psi_{i}^{(j)}(s) = \frac{\omega_{i,j}(s)s^{-1} + \alpha_{i,j}(s)}{\lambda_{j}(s)}$,\\
$\displaystyle\Psi_{i,1}^{(j)}(x,s) = \exp\left[\xi_{i,1}^{(j)}(s)(x-\ell_{i})\right]$ recalling $\ell_{m} = L$,\\
$\displaystyle\Psi_{i,2}^{(j)}(x,s) = \exp\left[\xi_{i,2}^{(j)}(s)(x-\ell_{i-1})\right]$ recalling $\ell_{0} = 0$,\\
$\displaystyle\xi_{i,1}^{(j)}(s) = \frac{v_{i} + \sqrt{v_{i}^{2} - 4D_{i}\lambda_{j}(s))}}{2D_{i}}$,\\
$\displaystyle\xi_{i,2}^{(j)}(s) = \frac{v_{i} - \sqrt{v_{i}^{2} - 4D_{i}\lambda_{j}(s))}}{2D_{i}}$.\\
\hline
\end{tabular}
\label{tab:FAB}
\end{table*} 

%\newpage
\section{Computational results}
\label{sec:results}
We demonstrate our generalized semi-analytical solutions derived in the previous sections using four test problems (Problems A--D). Problem A is a test case previously presented by \citet{sun_1999a} involving a homogeneous medium with a constant concentration of species 1 imposed at the inlet. Problem B, solved previously by \citet{suk_2013}, considers a heterogeneous three-layer medium with a constant concentration flux of species $1$ imposed at the inlet. Problem C involves a heterogeneous five-layer medium with a time-varying concentration of species $1$ imposed at the inlet and non-equal retardation factors across layers. Problem D involves constant zero order production of species $1$ in the fourth layer and distinct retardation factors for each species. Problems A and B consider sequential reactions with species 1 producing species 2, species 2 producing species 3 and species 3 producing species 4. Problem C considers a reaction network with species 1 producing species 2 and 3, species 2 producing species 3 and 4 and species 3 producing species 2 and 4. Problem D involves a similar reaction network to Problem C except species 1 also produces species 4 and species 4 produces species 2. Full information on each test problem is given in Table \ref{tab:test_problems}. 

\begin{table*}[p]
\centering\footnotesize\renewcommand{\arraycolsep}{2pt}
\caption{Transport and geometrical parameters for Problems A--D. In all problems, $c_{i,j}(x,t)$ has units of moles per litre [$\text{mol}\,\text{L}^{-1}$], initially the concentration is zero for all species across all layers, $c_{i,j}(x,0) = 0$, and a concentration gradient equal to zero is assumed at the outlet, $[\partial c_{m,j}/\partial x](L,t) = 0$ for all $j = 1,\hdots,n$. For Problems A--C, the zero-order production rate ($\gamma_{i,j}$) is zero for all $i = 1,\hdots,m$ and $j = 1,\hdots,n$. For Problem D, the zero-order production rate ($\gamma_{i,j}$) is equal to  $0.01\,\text{mol}\,\text{L}^{-1}\,\text{day}^{-1}$ for species 1 in layer 4 ($i = 4$ and $j = 1$) and zero otherwise.}
\label{tab:test_problems}

%Transport and geometrical parameters\\
\smallskip
\begin{tabular*}{1.0\textwidth}{@{\extracolsep{\fill}}lrrrrr}
\hline
 & $i$ & $\ell_{i}$ [$\text{m}$] & $D_{i}$ [$\text{m}^{2}\,\text{day}^{-1}$] & $v_{i}$ [$\text{m}\,\text{day}^{-1}$] & $\theta_{i}$\\
\hline
A & 1 & 50 & 0.3 & 0.2 & 1.0\\
  & 2 & 100 & 0.3 & 0.2 & 1.0\\
B & 1 & 0.3 & 0.01 & 1.0 & 0.12\\
  & 2 & 0.5 & 0.004 & 0.24 & 0.5\\
  & 3 & 1.0 & 0.01 & 1.0 & 0.12\\   
C--D & 1 & 10 & 0.08 & 0.4 & 0.09\\ 
  & 2 & 15 & 0.008 & 0.04 & 0.9\\
  & 3 & 20 & 0.08 & 0.4 & 0.09\\ 
  & 4 & 22 & 0.008 & 0.04 & 0.9\\
  & 5 & 40 & 0.08 & 0.4 & 0.09\\    
\hline
\end{tabular*}

\begin{tabular*}{1.0\textwidth}{@{\extracolsep{\fill}}llll}
\hline
 & First-order reaction [$\text{day}^{-1}$] & Retardation factors & Inlet boundary condition\\
\hline\\[0.00cm]
A & $\mathbf{M} = \displaystyle\left[\begin{matrix*}[r] -0.05 & 0 & 0 & 0\\ 0.05 & -0.02 & 0 & 0\\ 0 & 0.02 & -0.01 & 0\\ 0 & 0 & 0.01 & -0.005\end{matrix*}\right]$
& $R_{i,j}=1$. & $c_{1,j}(0,t) = \begin{cases} 1.0, & \text{$j = 1$,}\\ 0, & \text{$j = 2,3,4$.} \end{cases}$\\
&\\
B & $\mathbf{M} = \displaystyle\left[\begin{matrix*}[r] -0.75 & 0 & 0 & 0\\ 0.75 & -0.5 & 0 & 0\\ 0 & 0.5 & -0.2 & 0\\ 0 & 0 & 0.2 & -0.1\end{matrix*}\right]$ & $R_{i,j}=1$. & $v_{1}c_{1,j}(0,t) - D_{1}\dfrac{\partial c_{1,j}}{\partial x}(0,t) = \begin{cases} v_{1}, & \text{$j = 1$,}\\ 0, & \text{$j = 2,3,4$.} \end{cases}$\\
& \\
C & $\mathbf{M} = \displaystyle\left[\begin{matrix*}[r] -0.075 & 0 & 0 & 0\\ 0.05625 & -0.05 & 0.004 & 0\\ 0.01875 & 0.025 & -0.02 & 0\\ 0 & 0.025 & 0.016 & -0.045\end{matrix*}\right]$ & $R_{i,j} = \begin{cases} 1.0, & \text{$i = 1$,}\\ 0.8, & \text{$i = 2$,}\\ 1.0, & \text{$i = 3$,}\\ 0.8, & \text{$i = 4$,}\\ 1.0, & \text{$i = 5$.}\\ \end{cases}$ & $c_{1,j}(0,t) = \begin{cases} 1 - \exp(-0.01t), & \text{$j = 1$,}\\ 0, & \text{$j = 2,3,4$.} \end{cases}$\\
&\\
D & $\mathbf{M} = \displaystyle\left[\begin{matrix*}[r] -0.075 & 0 & 0 & 0\\ 0.045 & -0.05 & 0.004 & 0.045\\ 0.01875 & 0.025 & -0.02 & 0\\ 0.01125 & 0.025 & 0.016 & -0.045 \end{matrix*}\right]$ & $R_{i,j} = \begin{cases} 1.0, & \text{$j = 1$,}\\ 0.9, & \text{$j = 2$,}\\ 0.5, & \text{$j = 3$,}\\ 0.8, & \text{$j = 4$.}\\ \end{cases}$ & $c_{1,j}(0,t) = \begin{cases} 0.5(1+\cos(\pi t/800)), & \text{$j = 1$,}\\ 0, & \text{$j = 2,3,4$.} \end{cases}$\\
&\\
\hline
\end{tabular*}
\end{table*}
\begin{table}[p]
\centering\small
\caption{Maximum absolute difference between the semi-analytical and benchmark solutions shown in Figure \ref{fig:problems_AB} for test problems A--D (Table \ref{tab:test_problems}).}
\begin{tabular*}{0.8\textwidth}{@{\extracolsep{\fill}}lrrrr}
\hline
Problem & A & B & C & D\\
%Error & \num{9.1e-07} & \num{1.1e-06} & \num{3.3e-06} & \num{3.3e-06}\\ %5001
Error & \num{5.6e-07} & \num{5.5e-07} & \num{8.3e-07} & \num{1.0e-06}\\ %10001
%Error & \num{5.3e-07} & \num{5.8e-07} & \num{2.1e-07} & \num{9.8e-07}\\ %20001
\hline
\end{tabular*}
\label{tab:errors}
\end{table}

\begin{figure*}[p]
\centering
\includegraphics[width=\figw]{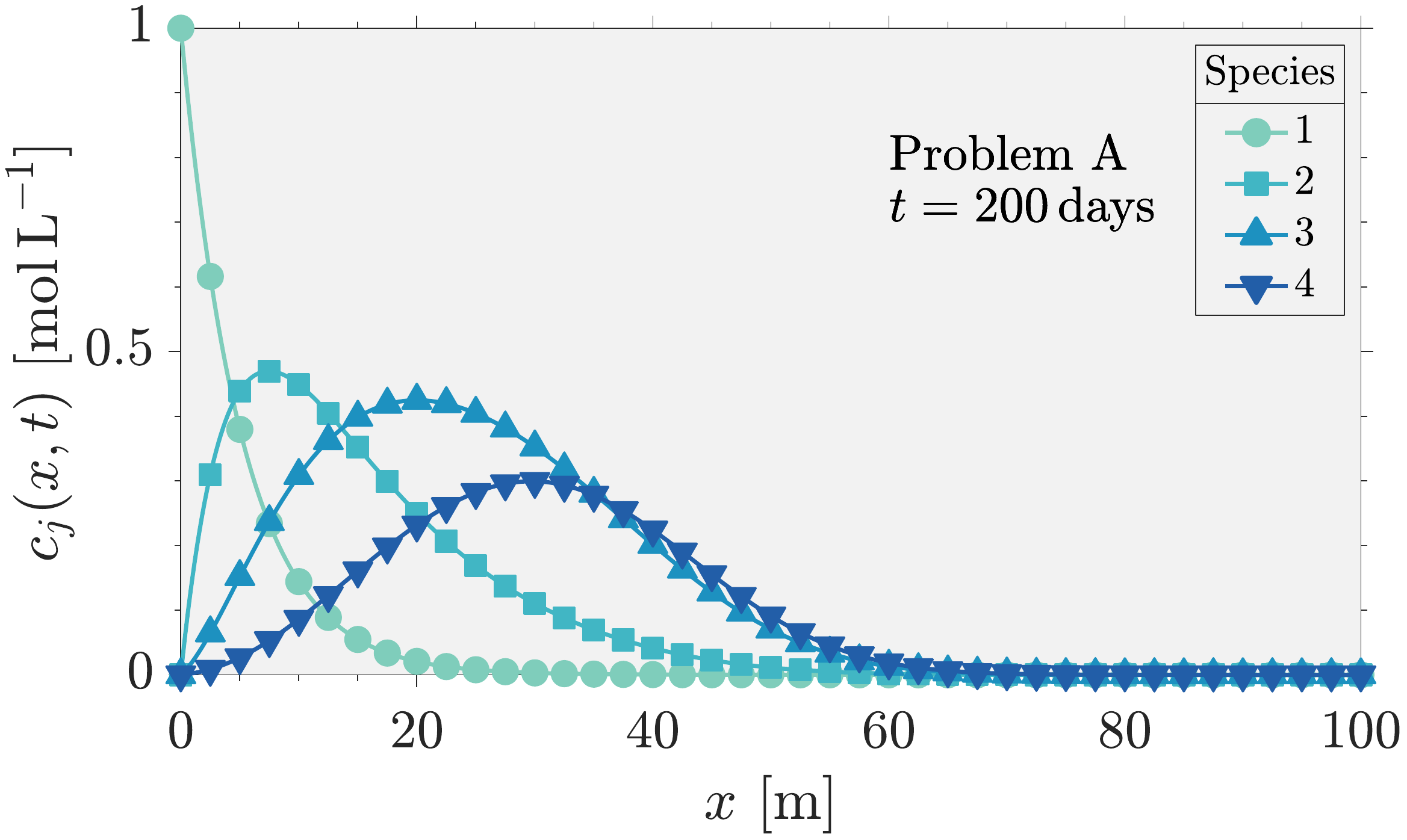}\hspace{0.05\textwidth}\includegraphics[width=\figw]{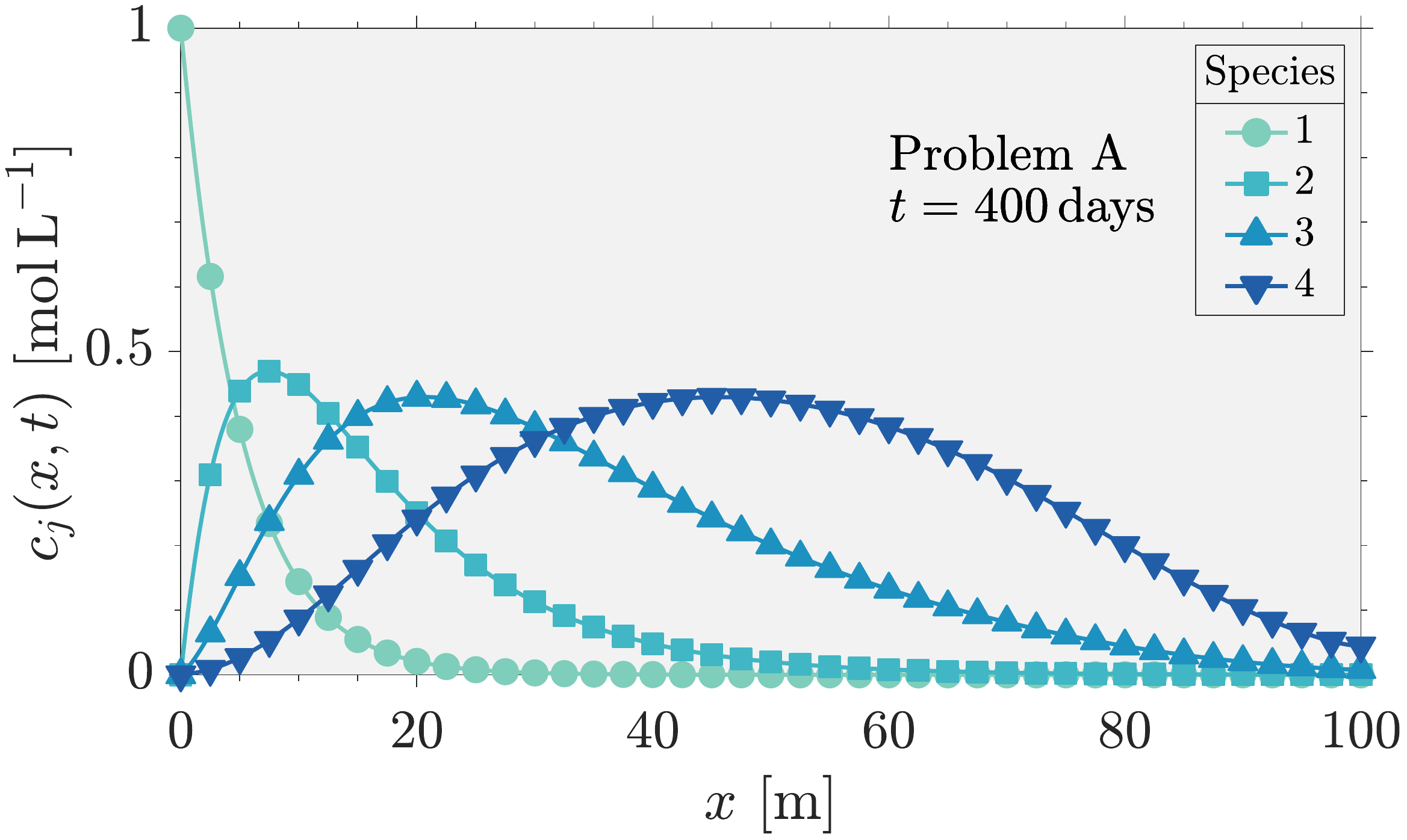}\\[0.3cm]
\includegraphics[width=\figw]{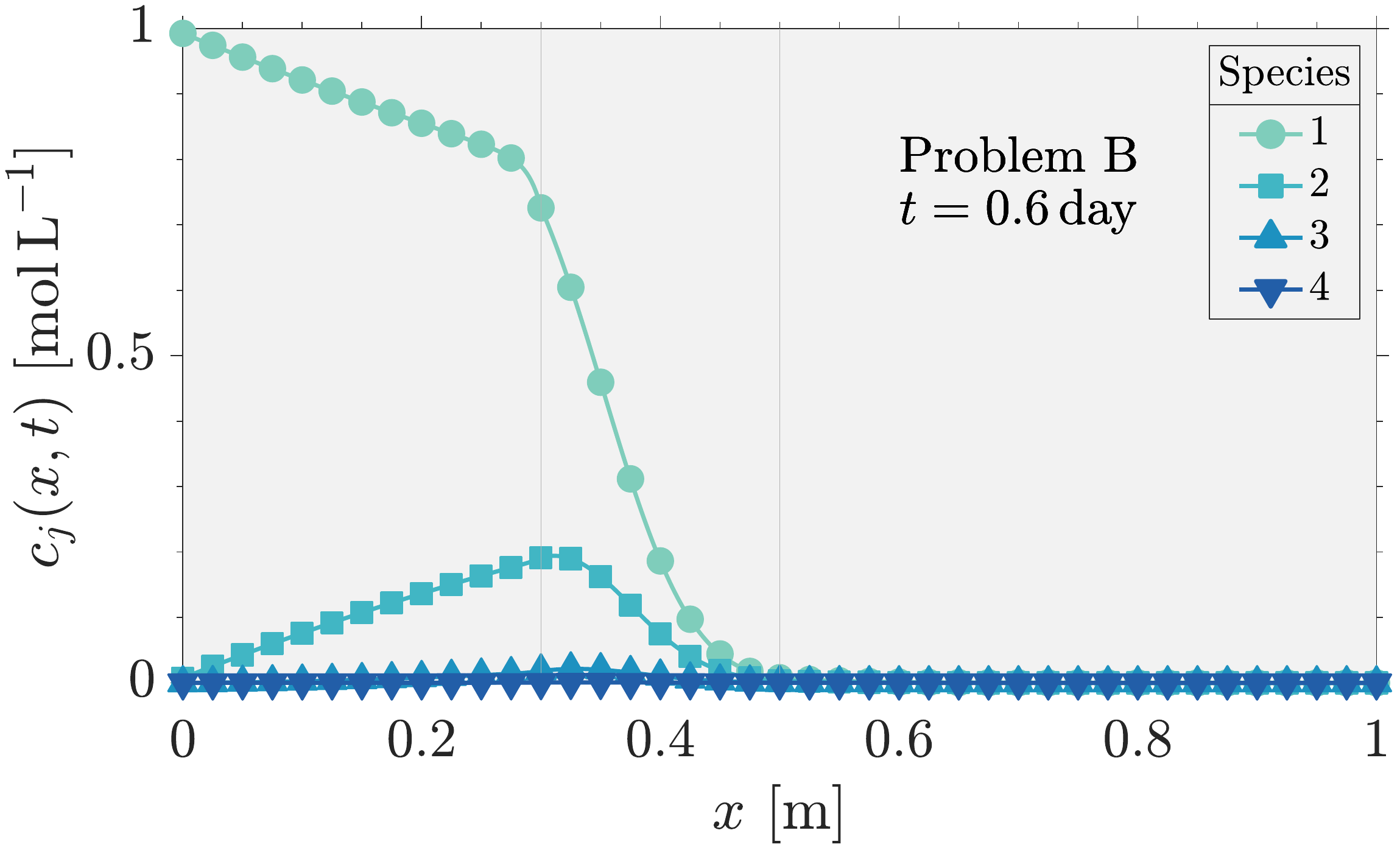}\hspace{0.05\textwidth}\includegraphics[width=\figw]{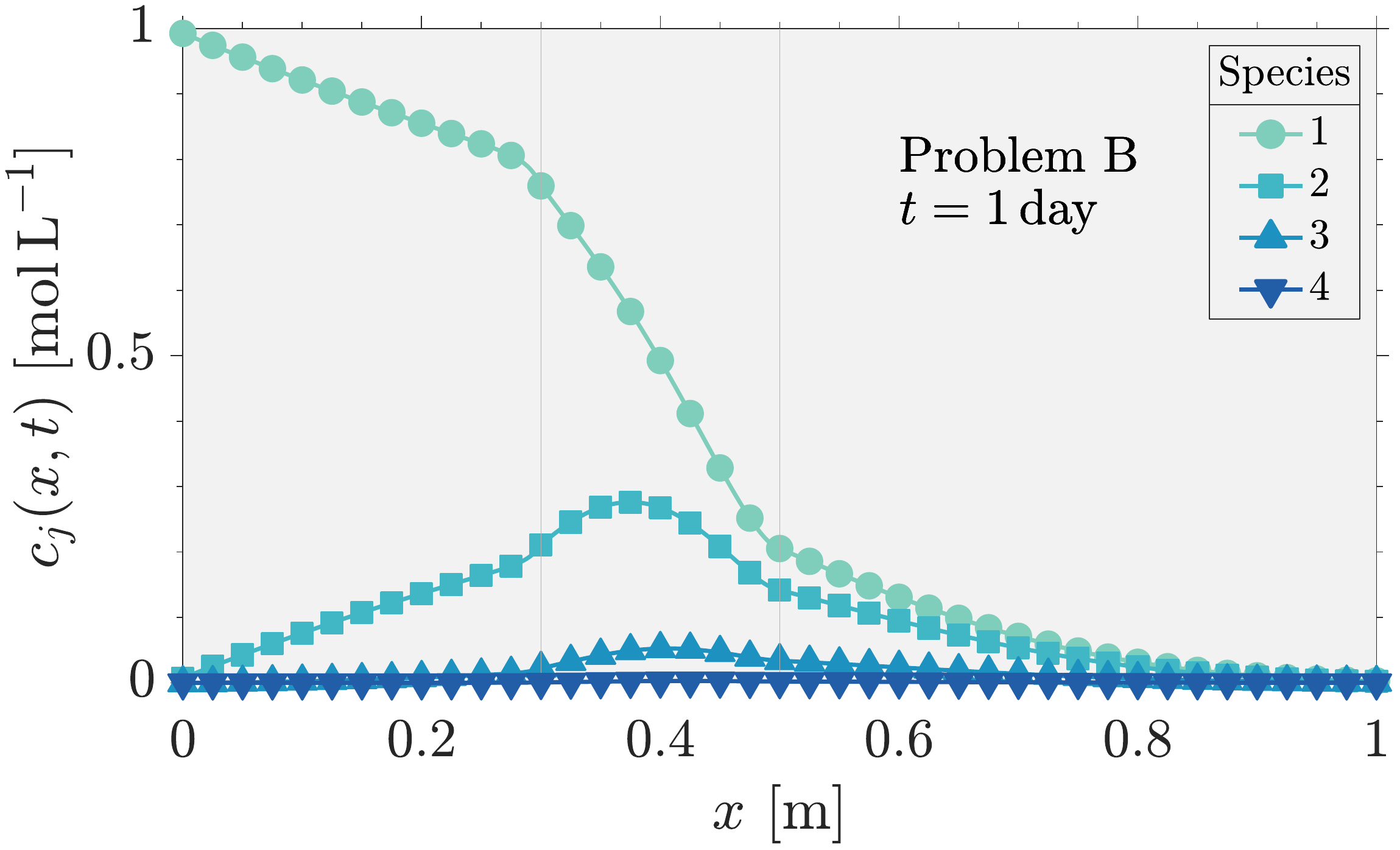}\\[0.3cm]
\includegraphics[width=\figw]{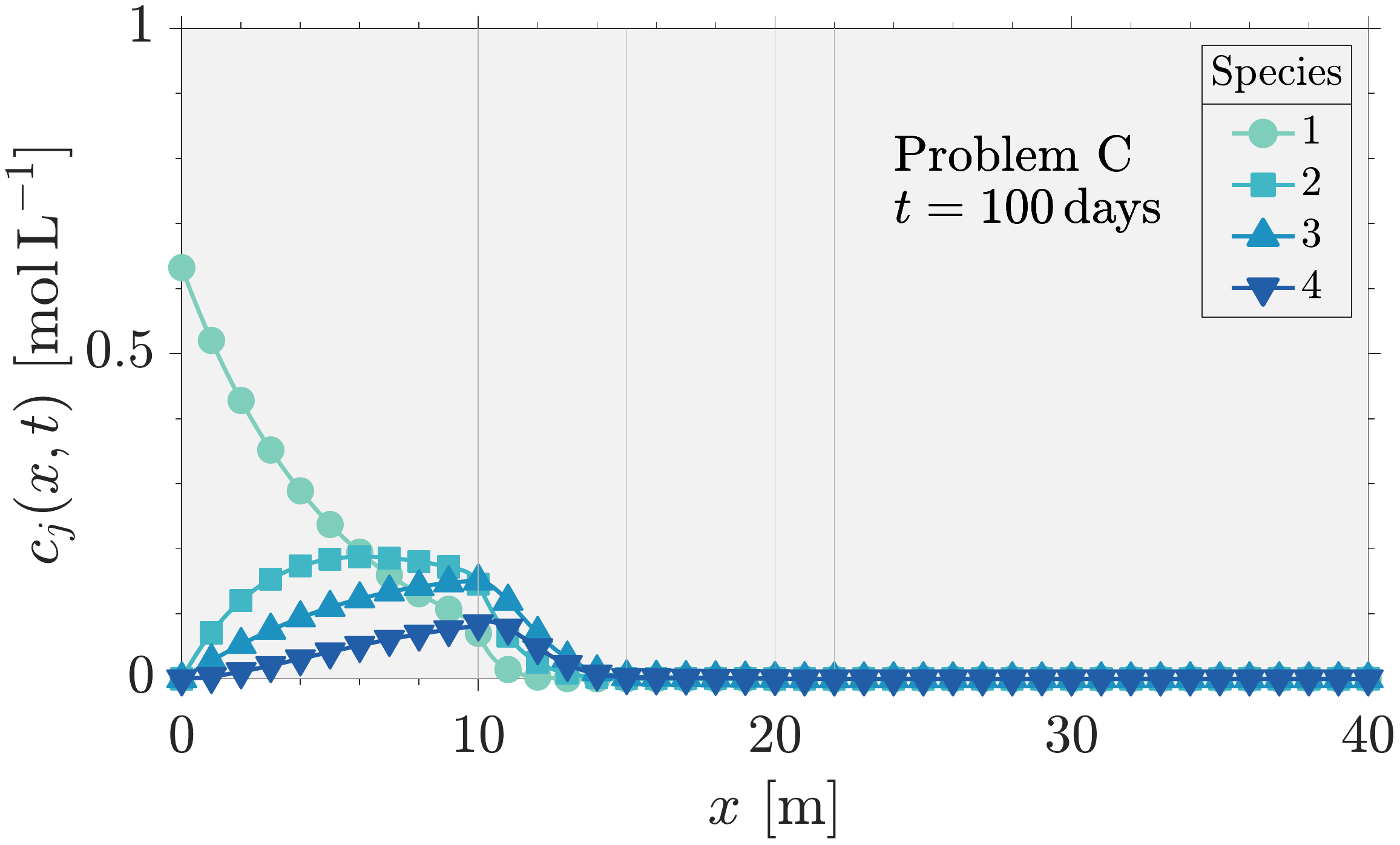}\hspace{0.05\textwidth}\includegraphics[width=\figw]{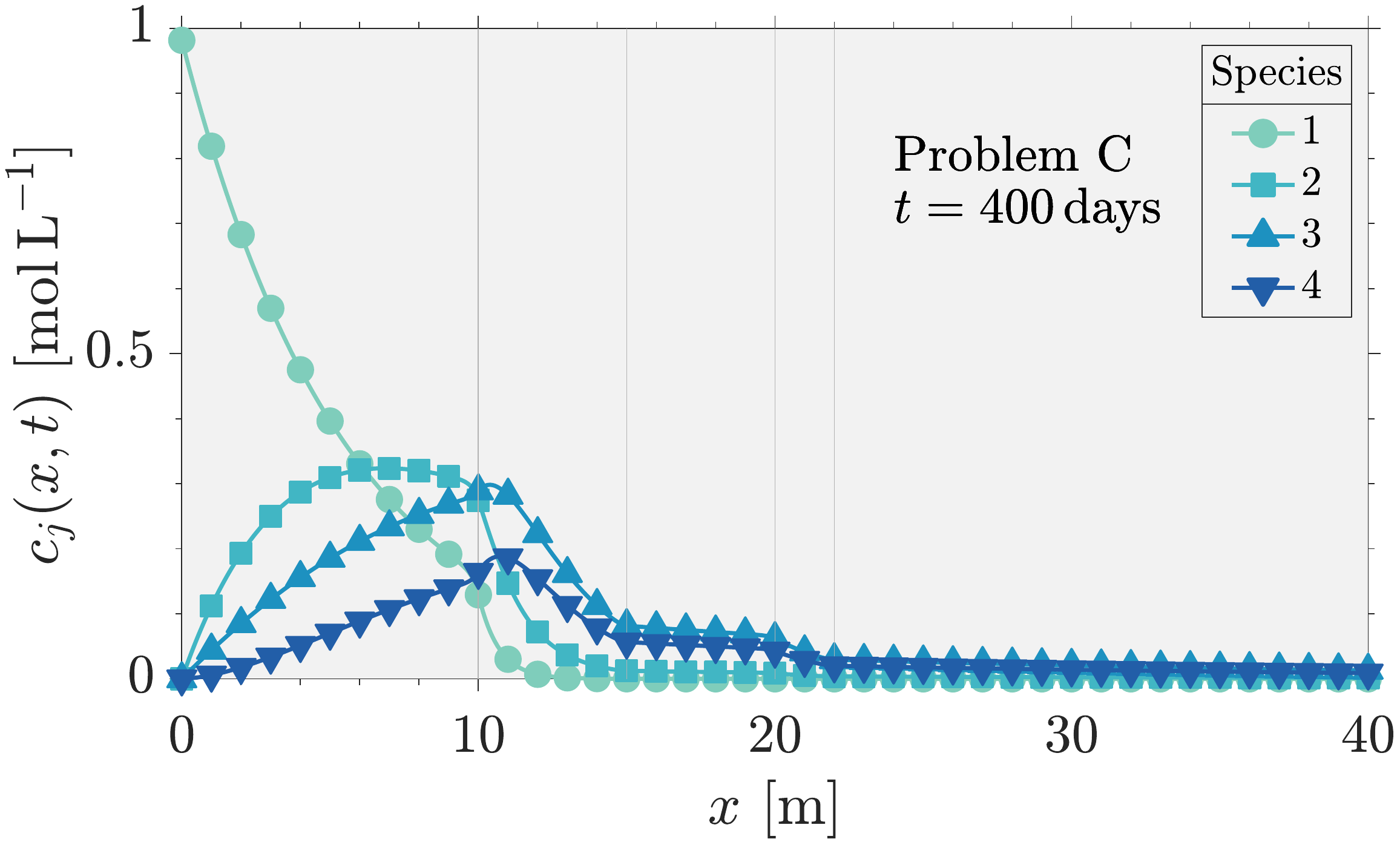}\\[0.3cm]
\includegraphics[width=\figw]{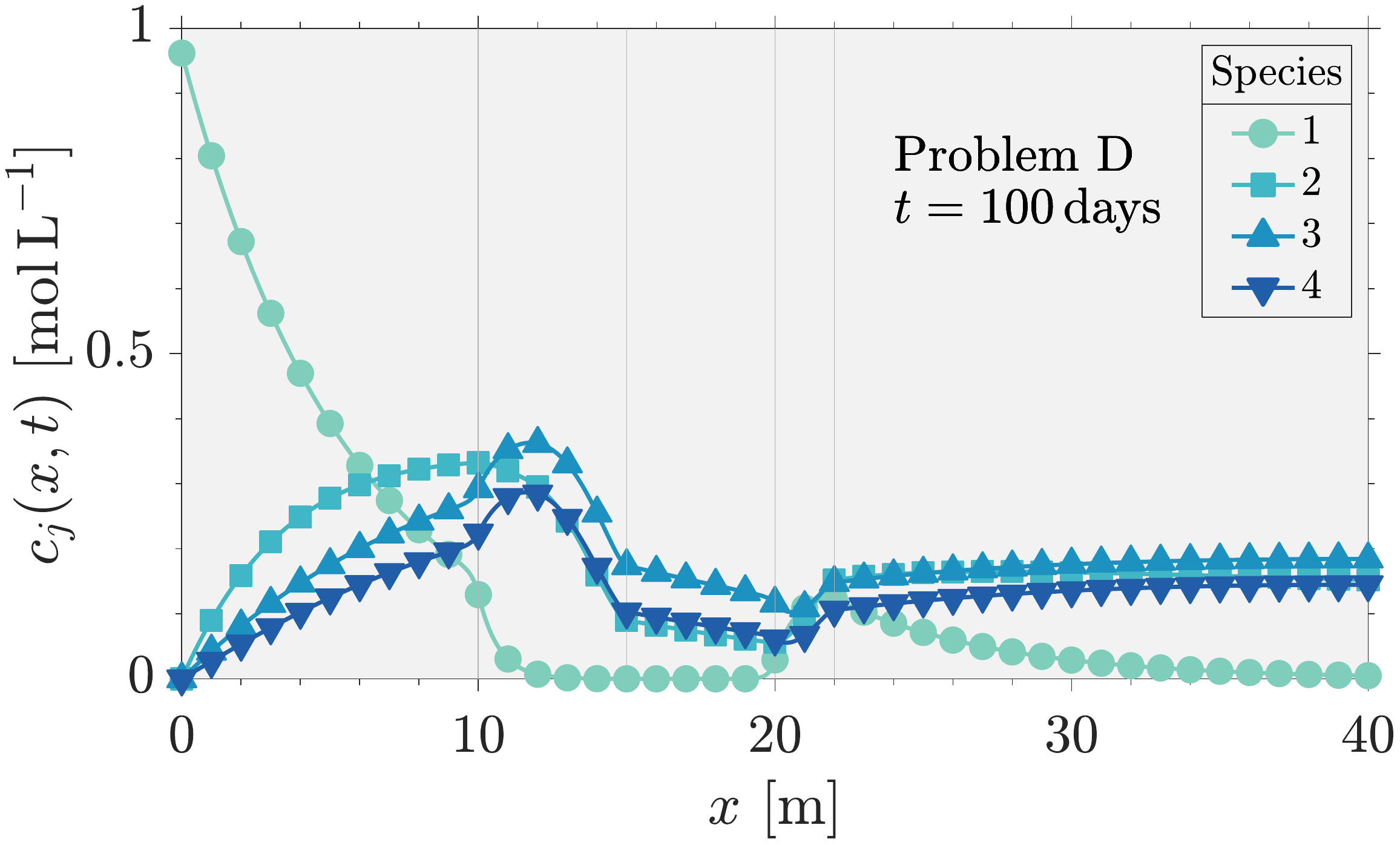}\hspace{0.05\textwidth}\includegraphics[width=\figw]{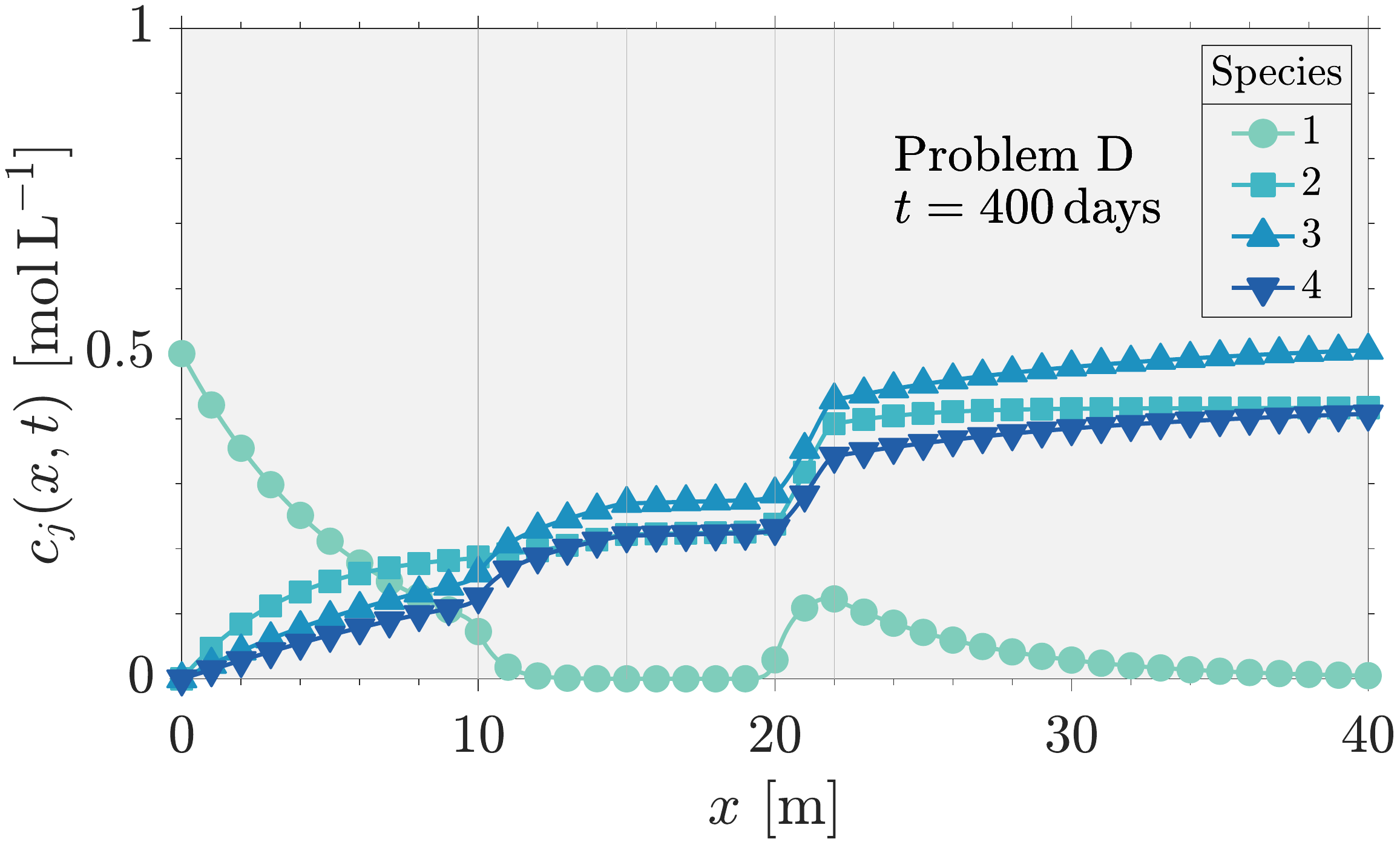}
\caption{Concentration profiles over time for Problems A--D where $c_{j}(x,t)$ denotes the concentration of species $j$ for all $x\in[0,L]$. In each plot, we compare our semi-analytical solution (solid line) to the benchmark numerical solution (markers) computed using $N = 10001$ nodes with markers shown at 41 (equally-spaced) points only. Interfaces between adjacent layers are represented using vertical lines.}
\label{fig:problems_AB}
\end{figure*}

\begin{figure*}[t]
\centering
\includegraphics[width=\figw]{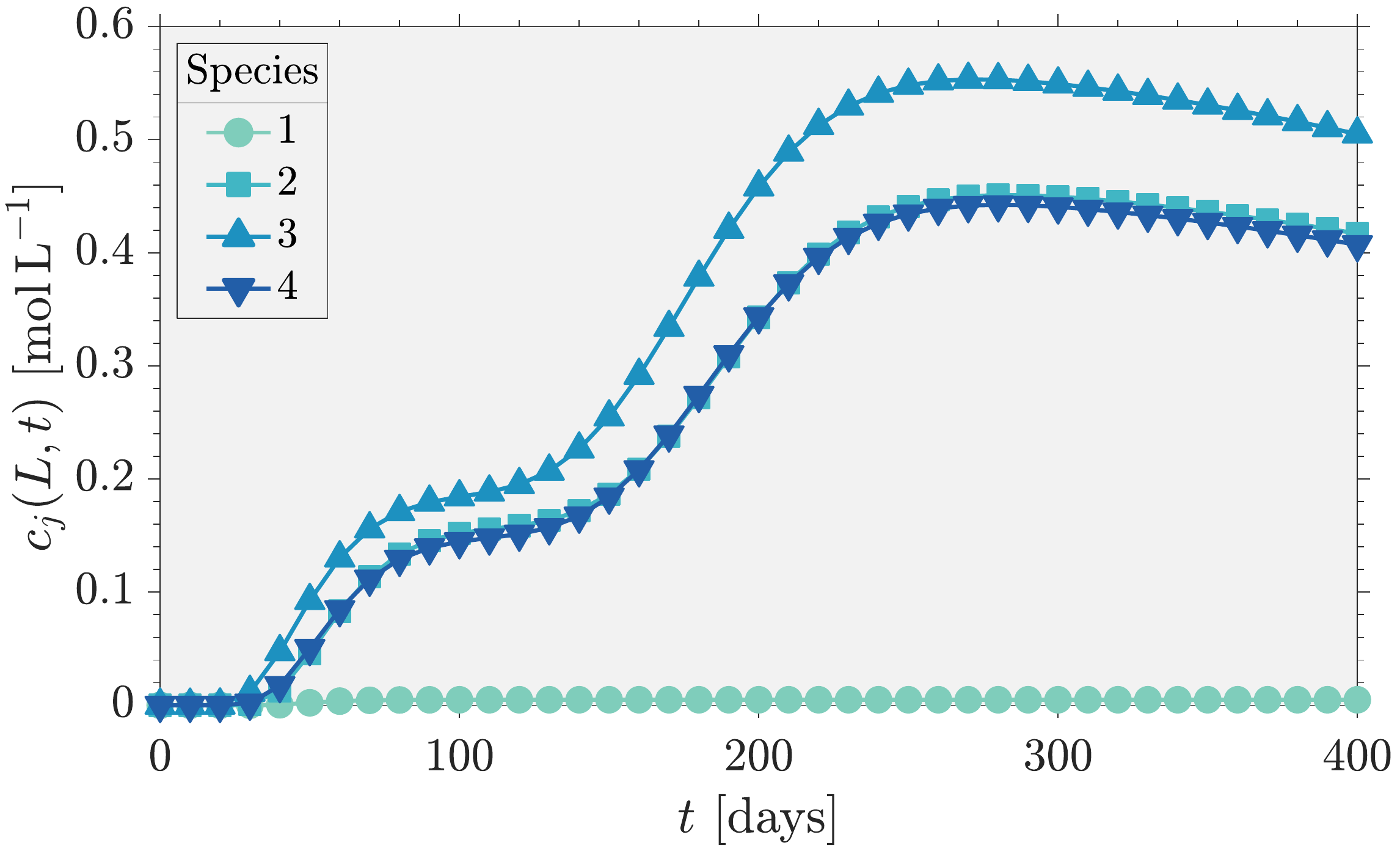}
\caption{Breakthrough curves for Problem D. The notation $c_{j}(x,t)$ denotes the concentration of species $j$ for all $x\in[0,L]$. In each plot, we compare our semi-analytical solution (solid line) to the benchmark numerical solution (markers) computed using $N = 10001$ with markers shown at $t = 0,10,20,\hdots,400\,\text{days}$.}
\label{fig:breakthrough}
\end{figure*}

For each test problem, we compare our generalized semi-analytical solution to a benchmark numerical solution. The benchmark solution is obtained by discretising the multispecies multilayer transport model (\ref{eq:c_pde1})--(\ref{eq:c_bc2}) in space using a standard finite volume method. Spatial discretisation is carried out on a fine uniform grid ($x_{i} = (i-1)L/(N-1)$ for $i = 1,\hdots,N$ and large $N$) with the resulting system of differential equations integrated in time using MATLAB's \texttt{ode15s} solver with strict error tolerances. Full implementation details of our numerical benchmark method, which generalises our single-species numerical method outlined in Appendix A of \citet{carr_2020} to multiple-species, is available from our MATLAB code (\href{https://github.com/elliotcarr/Carr2021a}{https://github.com/elliotcarr/Carr2021a}). Problems A--B involve equal retardation factors and Problem C non-equal retardation factors across layers, so we apply the Case I semi-analytical solution described in Section \ref{sec:distinct_layers} to solve these problems. On the other hand, we apply the Case II semi-analytical solution described in Section \ref{sec:distinct_species} for Problem D as it involves distinct retardation factors across species. For each test problem, results are presented in Figure \ref{fig:problems_AB} by plotting the concentration of each species across the medium at two points in time. In all cases, the benchmark solution with $N = 10001$ (markers) is visually indistinguishable from the semi-analytical solution (solid line). Additionally, the plots for Problems A--B match those presented previously by \citet{sun_1999a} and \citet{suk_2013}. In Table \ref{tab:errors}, we record the maximum absolute difference between the benchmark and semi-analytical solutions over time: $\text{Error} = \max |c_{i,j}(x_{k},t_{p}) - c_{i,j}(x_{k},t_{p})|$, where the maximums are taken over $i = 1,\hdots,m$ (layers), $j = 1,\hdots,n$ (species), $k = 1,\hdots,501$ ($x_{k} = (k-1)L/500$) and $p = 1,\hdots,2$ ($t_{1}$ and $t_{2}$ are the discrete times shown in Figure \ref{fig:problems_AB}). These results reveal that the semi-analytical solutions for Problems A--D agree with the benchmark finite volume solution ($N = 10001$ nodes), to at least five decimal places. 

As mentioned in the introduction, a clear advantage of analytical and semi-analytical solutions is that they are continuous in space and time. Our generalized semi-analytical solution can therefore be evaluated at a single point in space or time without affecting accuracy. In contrast, numerical solutions require sufficiently small spatial and temporal discretisation step sizes to ensure accurate computation of solutions at a single point in space or time. Figure \ref{fig:breakthrough} profiles the breakthrough curves (species concentration at the outlet) for Problem D, computed using the generalized semi-analytical solution and the numerical benchmark solution with $N = 10001$. Generating the breakthrough curves using our semi-analytical solution takes less than one twentieth of the time required by the numerical benchmark solution.

\section{Conclusion}
\label{sec:conclusion}
In this paper, we developed a generalized semi-analytical solution for multispecies multilayer advection-dispersion equations coupled via first-order reactions. Our solution strategy combines a transformation for decoupling multispecies equations \cite{clement_2001,quezada_2004} with our previous semi-analytical method for solving single-species multilayer advection-dispersion-reaction problems \cite{carr_2020}. The derived solutions are valid for arbitrary numbers of species and layers, general first-order reaction networks and general Robin-type conditions at the boundaries (inlet and outlet). Seperate solutions were presented for accommodating (i) non-equal retardation factors across layers (Case I, Section \ref{sec:distinct_layers}) and (ii) non-equal retardation factors across species (Case II, Section \ref{sec:distinct_species}). Our work addresses limitations of existing analytical or semi-analytical solutions that are restricted to homogeneous media, specific reaction networks, certain choices of boundary conditions or a combination thereof. \revision{Our semi-analytical solution broadens the suite of such solutions for reactive contaminant transport problems, allowing application of semi-analytical solutions to a broader range of important problems involving general first order reaction networks in multilayer media such as contaminant transport through composite liners used in landfill containment systems \cite{chen_2009}.}

While our solution is very general, it is limited in some aspects. For example, for the case of non-equal retardation factors across layers (Case I, Section \ref{sec:distinct_layers}), the reaction matrix $\mathbf{M}$ must be diagonalisable and cannot vary across layers, the coefficients $a_{0}$, $b_{0}$, $a_{L}$ and $b_{L}$ appearing in the inlet and outlet boundary conditions must be the same across species and the retardation factors cannot vary across both layers and species. If any of these do not hold, the linear transformation methods (outlined in Section \ref{sec:solution_method}) fail to remove the coupling between species. Additionally, we found that the algorithm employed for numerically inverting the Laplace transform can lead to unreliable results for advection-dominated problems. Possible directions for future work include addressing some of these limitations. 

\section*{Acknowledgements}
\noindent Funding for this research was provided by the Australian Research Council (DE150101137).

%\section*{References}
\bibliographystyle{elsarticle-num-names}
\bibliography{references}

\end{document}